\documentclass[aps,prl,onecolumn,superscriptaddress,
floats,floatfix,preprintnumbers,longbibliography,subfigure,nofootinbib]{
revtex4}

\usepackage{amssymb,amsmath}
\usepackage{bm,url,graphics,subfigure,epsfig}
\usepackage{ulem}
\usepackage{color}
\usepackage{tikz}
\usepackage{pdfpages}

\def\ADD#1{{\textcolor{black}{#1}}}         

\def \bE {\mathbf{E}}
\def \nE {\bar{E}}

\def \be {\mathbf{e}}

\def \br {\mathbf{r}}

\def \bN {\mathbf{N}}
\def \bQ {\mathbf{Q}}

\def \bez {\be_z}

\def \ortp {\be_\mathrm{p}}
\def \d {\mathrm{d}}

\def \nonOmega {\bar{\Omega}}
\def \nont {\bar{t}}
\def \nonOmegapk {\nonOmega^{\mathrm{lpk}}}
\def \nonOmegaamp {\nonOmega^{\mathrm{mag}}}
\def \bareta {\bar{\eta}}

\def \barmu {\bar{\mu}}
\def \barmuopt {\barmu^{\mathrm{opt}}}
\def \barmucrione {\barmu_1^{\mathrm{cri}}}
\def \barmucritwo {\barmu_2^{\mathrm{cri}}}

\def \Utran {\bar{\mathcal{U}}}
\def \Utranopt {\bar{\mathcal{U}}^{\mathrm{opt}}}

\def \barE {\bar{E}}
\def \barEcrione {\barE_1^{\mathrm{cri}}}
\def \barEcritwo {\barE_2^{\mathrm{cri}}}

\def \bP {\boldsymbol{\mathcal{P}}}

\def \br {\mathbf{r}}

\def \esl {\epsilon_{\mathrm{sl}}}

\def \Ecri {E^{\mathrm{cri}}}

\def \lp {\left(}
\def \rp {\right)}

\def \sp {\sigma_{\mathrm{p}}}
\def \ss {\sigma_{\mathrm{s}}}
\def \ep {\epsilon_{\mathrm{p}}}
\def \es {\epsilon_{\mathrm{s}}}

\def \taus  {\tau_{\mathrm{s}}}

\def \nP {\bar{\mathcal{P}}}

\def \nTfp {\bar{\Gamma}^{\mathrm{f}\rightarrow\mathrm{p}}}

\begin{document}

\title{Propulsion driven by self-oscillation via 
an electrohydrodynamic instability}\vspace{-2ex}

 \author{Lailai Zhu}
 \affiliation{Department of Mechanical and Aerospace Engineering, Princeton 
University, Princeton, NJ 08544, USA}
 \affiliation{Linn\'{e} Flow Centre and Swedish e-Science Research Centre (SeRC), KTH 
 Mechanics, Stockholm, SE-10044, Sweden}
 \author{Howard A. Stone}
 \email{hastone@princeton.edu}
 \affiliation{Department of Mechanical and Aerospace Engineering, Princeton 
University, Princeton, NJ 08544, USA}

\begin{abstract}
Oscillations of flagella and cilia play an important role in biology,
which motivates the idea of functional mimicry as part of 
bio-inspired applications. 
Nevertheless, it still remains challenging to drive their artificial 
counterparts to oscillate 
via a steady, homogeneous stimulus.
Combining theory and simulations, we demonstrate a strategy to achieve 
this 
goal by
 using an elasto-electro-hydrodynamic instability.
In particular, we show that applying a uniform DC electric field can produce 
self-oscillatory
motion of a microrobot composed of a dielectric particle and an elastic 
filament.
Upon tuning the electric field and filament elasticity, the 
microrobot exhibits three distinct behaviors: a stationary state, undulatory 
swimming and steady spinning, where the swimming behavior stems from an 
instability emerging through a Hopf bifurcation. Our results imply the 
feasibility of
engineering self-oscillations by leveraging the 
elasto-viscous response to control the type of 
bifurcation and the form of instability. We 
anticipate that 
our strategy will be useful
 in a broad range of applications imitating self-oscillatory natural
phenomena and biological processes.
\end{abstract}

\maketitle

Flagella and cilia exhibit oscillatory movements for 
locomotion, pumping and fluid mixing. To mimic these functionalities, 
various approaches have been developed to oscillate their 
artificial counterparts using
magnetic~\cite{singh2005synthesis,evans2007magnetically},
 electrostatic~\cite{den2008artificial}, piezoelectric~\cite{oh2009bio},  
 optical~\cite{van2009printed} and 
hydrogel-based actuations~\cite{sidorenko2007reversible,masuda2013self}.
In general, a time-dependent 
stimulus generates 
oscillations in many biomimetic systems. An apparent exception 
is the use of the Belousov-Zhabotinsky (BZ) oscillating chemical
reaction~\cite{masuda2013self} (inspired by 
Ref~\cite{yoshida1996self}) to deform polymer brushes periodically.
Nonetheless, time-dependent forcing is not necessary for biological 
systems which can occasionally generate 
oscillations by steady stimuli to deliver functionalities such as  
otoacoustic emissions~\cite{gold1948hearing,kemp1979evidence} and 
glycolysis~\cite{sel1968self}, etc. 
These behaviors, namely, the 
generation and maintenance of a periodic motion powered by a source without a 
corresponding periodicity is referred 
to as self-oscillation~\cite{jenkins2013self,sc}.

Self-oscillation plays a crucial role in some inertia-dominated  flow 
cases, such as the collapse of the 
Tacoma Narrows Bridge~\cite{bridge} and the sound generation of wind 
musical instruments (including whistling and the human voice), owing to 
inertia-induced nonlinearity. \ADD{In this Rapid Communication}, we create 
self-oscillations of 
artificial structures in
a situation with negligible inertia by applying a uniform, time-independent 
electric field. We 
exploit an elasto-electro-hydrodynamic (EEH) instability by 
marrying an electrohydrodynamic instability 
with an elasto-viscous response. Combining theory and simulations,
we investigate a composite 
microrobot that achieves unidirectional locomotion by self-oscillatory 
wiggling of an elastic appendage.


%
%

It was discovered in 1896 by 
Quincke~\cite{quincke1896ueber} that a uniform DC electric field can trigger 
the spontaneous rotation of a dielectric particle immersed in 
a dielectric solvent with higher conductivity. Quincke rotation (QR)
occurs as an electrohydrodynamic instability emerging from an  equilibrium 
configuration where
the induced-charge dipole $\bP$ of the particle is opposite to the 
applied field. When 
the field strength $E$ exceeds
 a threshold value, $\Ecri$, the symmetric yet antiparallel
  configuration is unstable to an infinitesimal 
disturbance, spontaneously breaking the mirror symmetry through a 
supercritical 
pitchfork bifurcation~\cite{turcu1987electric,peters2005experimental}; the 
particle then spins steadily where the electric and viscous torques 
balance. 

\begin{figure*}[!ht]
\begin{center}
\includegraphics[width=0.95\textwidth]
{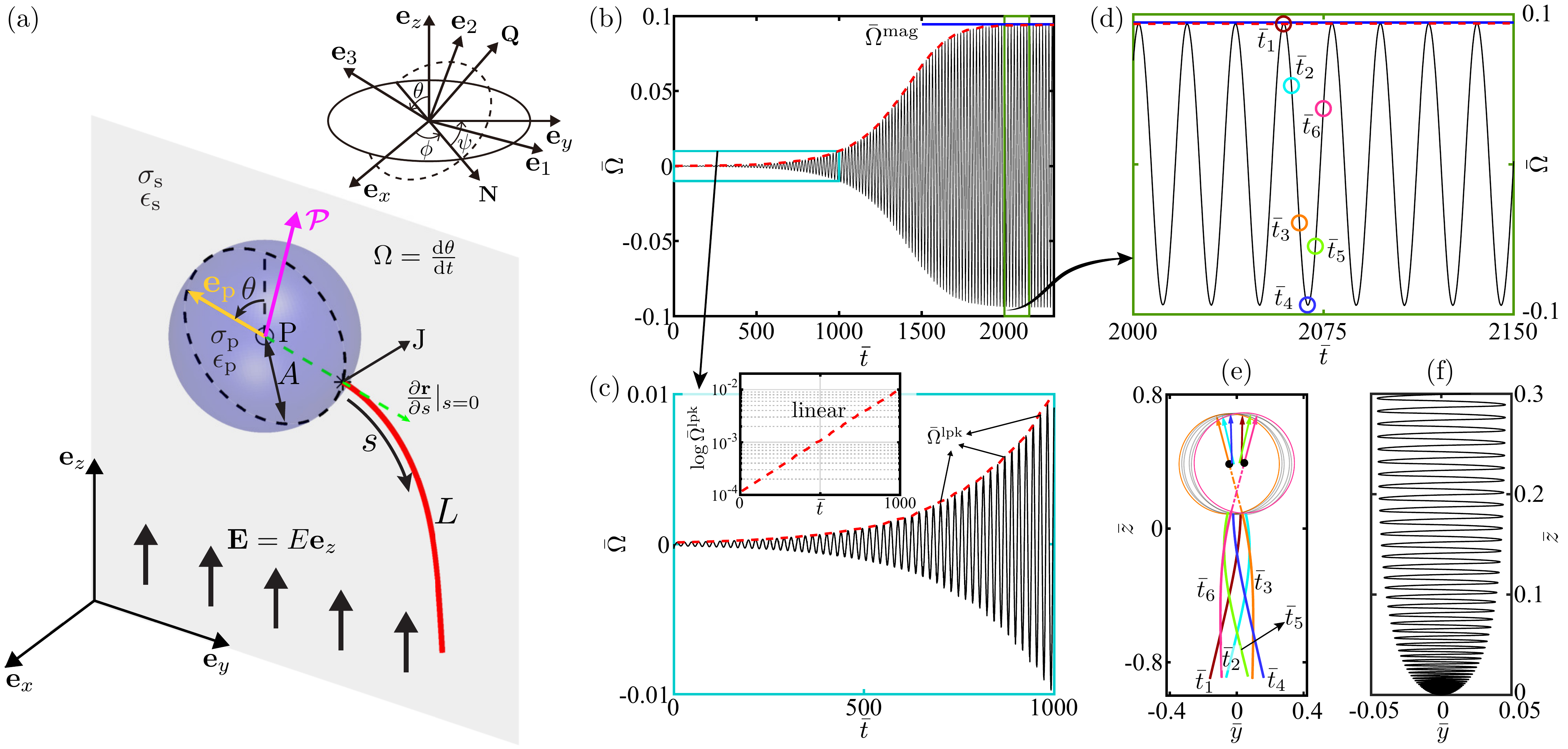}
\end{center}
\caption{Self-oscillation of a composite particle and filament.
(a) A dielectric sphere of radius $A$, with an attached filament of 
radius $a$, contour length $L$, size ratio $\alpha=A/L=0.3$ and elasticity
characterized by the EEV number $\barmu=635$, is
under a steady uniform electrical field $\bE = E\bez$; the 
particle dipole is $\bP$.  The upright panel depicts
the Euler angles $[\theta,\phi,\psi]$ describing the 
rotation of the object and the reference coordinate 
system $[\be_N, \be_Q, \be_3]$, where 
$\be_3$ is aligned with the object's orientation $\ortp$. The 
object's motion is constrained to the 
$yz$-plane, and its orientation $\ortp \equiv \be_3$  is directed 
at an angle $\theta$ with respect to $\be_z$.
(b) The composite object's 
dimensionless 
rotational velocity $\nonOmega = \d\theta/ \d \nont$ oscillates over 
dimensionless time $\nont$ with a local peak value $\nonOmegapk$. The time 
evolution of $\nonOmega$ is 
characterized by two periods, as $\nonOmegapk$ increases exponentially in the 
initial phase (cyan box) and 
eventually reaches a  constant magnitude $\nonOmegaamp$ (green box).
 (c) Exponential growth of $\nonOmega$ in the initial 
phase $\nont \in [0, 1000]$ is   indicated by the inset showing the linear 
dependence of $\log{\nonOmegapk}$ 
on $\nont$. (d) Periodic variation of $\nonOmega$ with a constant 
amplitude; six times 
 $\nont_i, i=1,..,6$   within one period are marked. (e) Orientation of 
the composite object
and the filament shape at $\nont_i$. (f) Trajectory of the particle 
center P in the $yz$-plane  during the period $\nont\in [0, 1940]$.}
\label{fig:fig1}
\end{figure*} 

We exploit this QR instability by grafting an inextensible 
elastic filament of radius $a$ and length $L$
onto a dielectric spherical particle of radius $A$ (Fig.~\ref{fig:fig1}a), where 
$s$ denotes the arclength of the filament's centerline with position 
$\br(s,t)$, and $\alpha=A/L$ is the size ratio. The slenderness of 
the filament $\esl = a/L  \ll 
1$. The filament base $s=0$ is clamped at the 
particle surface J, where the tangent vector 
$\ortp = \partial\br/\partial 
s|_{s=0}$ at the base
always passes through the particle center P and $\ortp$ denotes the orientation
of the object.
We define an 
elasto-electro-viscous (EEV) parameter $\barmu = 8\pi \mu 
L^4/D\taus$ 
indicating the ratio of the elasto-viscous time scale $8\pi \mu L^4/D$  to
the charge relaxation time $\taus=\es/\ss$
of the solvent, 
where  $\mu$, $\es$ and $\ss$ denote respectively, the dynamic viscosity, 
permittivity and conductivity  of the solvent,  and $D$ is the 
bending stiffness of the filament.
$\barmu$ indicates the relative strength of the viscous to the 
elastic forces, where $\barmu=0$ corresponds to 
a rigid filament and increasing $\barmu$ corresponds to a more compliant 
filament. To focus on the elasto-viscous response of the filament, we do not 
consider its polarization. \ADD{We also do not take into account the 
hydrodynamic
interactions between the particle and the filament.}

We adopt the \ADD{proper} Euler angles $[\theta,\phi,\psi]$ to characterize
the rotation of the object. The uniform electric field $\bE = E \be_z$
and the particle dipole $\bP$  are expressed in the 
reference coordinate system $[\bN, \bQ, 
\be_3]$ rotating and translating with the object, where $\be_3$ coincides 
with the its orientation $\ortp$,  $\bN$ indicates the 
nodal line direction and $\bQ = \be_3 \times \bN$ (Fig.~\ref{fig:fig1}a). We 
constrain the object's motion to 
the $yz$-plane, hence the  dipole $\bP$, the orientation $\ortp$, and the 
filament lie in the same plane, resulting in $\bN =\be_x$. Using $\taus$, 
$\taus^{-1}$, $D/L$,
$\Ecri$ and $D/\lp L \Ecri \rp$ as the characteristic time, rotation rate, 
torque, electrical field and polarization dipole strength, 
respectively, the 
nondimensional electrohydrodynamic 
equations are~\cite{cebers2000electrohydrodynamic} (see Supplemental 
Material)
\begin{subequations}\label{eq:non-euler}
\begin{align}
 \partial \theta/\partial \nont & = \lp\nTfp_N + 
\nE_3 \nP_Q 
- \nE_Q \nP_3 \rp/\bareta,  \label{eq:non-theta} \\
\partial \nP_Q/\partial \nont & = -\kappa \lp\nP_Q + \kappa \bareta 
\nE_Q \rp,  \label{eq:non-PQ} \\
\partial \nP_3/\partial \nont & = -\kappa \lp \nP_3 + \kappa \bareta 
\nE_3 \rp, \label{eq:non-P3}
\end{align} 
\end{subequations}
where $\;\bar{}\;$ denotes dimensionless quantities hereinafter,
$\bar{\boldsymbol{\Gamma}}^{\mathrm{f}\rightarrow\mathrm{p}}$ is the elastic 
torque exerted by the filament onto the particle with respect to its center;
$\bareta = 
\alpha^3 \barmu$, $\kappa =(R+2)/(S+2)$, $R=\sp/\ss$ and $S=\ep/\es$,
where $\ep$ and $\sp$ are the 
permittivity and conductivity of the particle, respectively.

\begin{figure*}[!ht]
\begin{center}
\includegraphics[width=0.85\textwidth]
{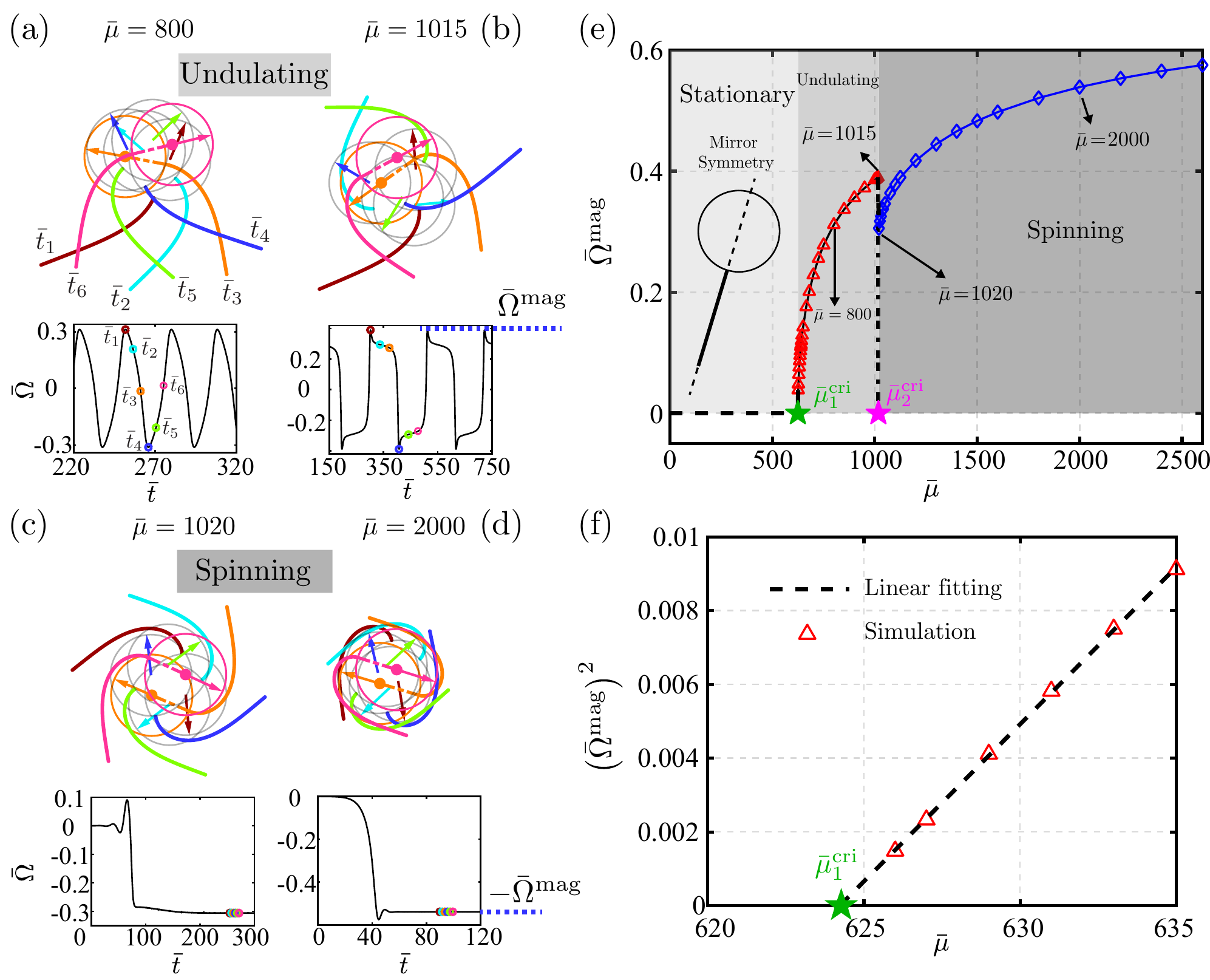}
\end{center}
\caption{Tuning the filament elasticity for various behaviors: 
stationary, undulating and spinning.
(a) and (b) Results for $\barmu=800$ 
and $1015$: 
time evolution of $\nonOmega$ when the object undulates periodically 
with a magnitude $\nonOmegaamp$, and the particle-filament
configuration at six times within a period. The $\barmu=1015$ 
case exhibits a large amplitude in the rotational displacement.
(c) and (d) Similar to (a) and 
(b), but for $\barmu=1020$ and $2000$,
when the system evolves from $\nont=0$ to its steady spinning state; six 
times are chosen within one rotational period. (e) The magnitude of 
rotational velocity $\nonOmegaamp$ versus $\barmu$. 
The system is stationary when $\barmu < \barmucrione \approx 625$, undulates
 when $ \barmucrione < \barmu < \barmucritwo \approx 1017$
and spins steadily when $\barmu > \barmucritwo$. Two undulatory cases 
($\barmu=800$ and $1015$) and two steady spinning ones ($\barmu=1020$ and 
$2000$)
are marked. (f) Linear relation between $(\nonOmegaamp)^2$ and $\mu$ close 
to 
$\barmucrione$. }
\label{fig:fig2}
\end{figure*} 

Applying a fixed electric field, $\nE = 
1.5$ for example, we 
discovered a self-oscillatory response of the composite 
object by tuning the filament elasticity (Fig.~\ref{fig:fig1}).
When $\barmu=635$, the particle  
wobbles spontaneously rather than to rotate steadily 
like the classical QR counterpart, as 
indicated by the  
rotational velocity $\nonOmega = \d \theta /\d \nont$
(Fig.~\ref{fig:fig1}b). 
The local peak $\nonOmegapk$ of $\nonOmega$
increases with $\nont$ rapidly during the initial period and eventually 
saturates to a constant value $\nonOmegaamp$ corresponding to 
a time-periodic state. 

To understand the initial dynamics,
we examine the initial phase $\nonOmega 
(\nont)$, as shown in Fig.~\ref{fig:fig1}b (highlighted in the cyan box). 
This local peak $\nonOmegapk$  
(Fig.~\ref{fig:fig1}c) initially grows exponentially, as
confirmed by the inset displaying the linear dependence of 
$\log{\nonOmegapk}$ on $\nont$. Thus,   
the self-oscillation arises through a linear 
instability mechanism, similar to other self-oscillation 
phenomena~\cite{jenkins2013self}.
Furthermore, the system reaches a time-periodic state, namely,
the particle oscillates with a fixed amplitude (Fig.~\ref{fig:fig1}d 
highlighting
 the green box of Fig.~\ref{fig:fig1}b). 
To understand how the grafted filament reacts to the particle,
we show in Fig.~\ref{fig:fig1}e the particle-filament 
configurations at six 
times within a period. We observe that the oscillating particle drives
the filament to wiggle, a scenario resembling the locomotion of a flagellated 
microorganism that acquires thrust by
propagating oscillatory bending waves from the head towards the tail. 
A striking yet natural consequence of this self-oscillation is that the
object undulates and translates, hence demonstrates Propulsion by 
harnessing thrust from the wiggling filament.

\begin{figure*}[!ht]
\begin{center}
\includegraphics[width=0.9\textwidth]
{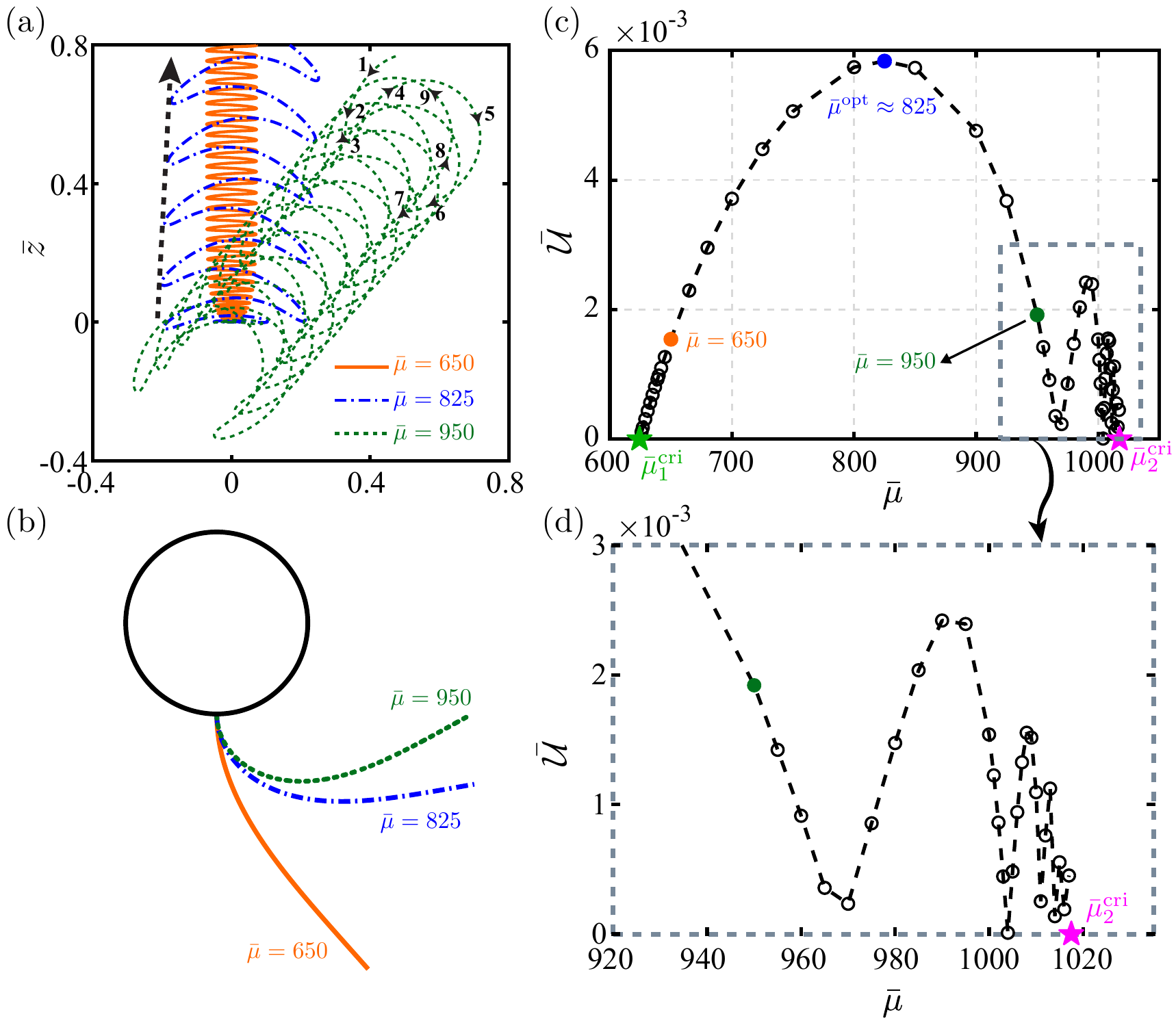}
\end{center}
\caption{Self-oscillatory propulsion of the composite swimmer. 
(a) Planer swimming trajectories for $\barmu=650$, $825$ 
 and $950$. The dashed arrow
 indicates the on average straight trajectory of the $\barmu=825$ 
swimmer. Arrows 
with numbers are marked at increasing times on the trajectory of $\barmu=950$.
(b) Maximum filament defection. 
(c) Effective  translational velocity $\Utran$ versus 
$\barmu$, reaching its peak  at 
$\barmu = \barmuopt \approx 825$. \ADD{(d) Close-up view of $\Utran(\barmu)$ in 
the regime 
where $\Utran$ exhibits a wavelike variation near $\barmu \approx 
\barmucritwo$.}}
\label{fig:fig3}
\end{figure*}

A series of  simulations was performed to examine the influence of 
the filament elasticity. By varying $\barmu$,
we identify three states of the composite object:
an undulatory motion (Fig.~\ref{fig:fig2}a,b, where $\barmu=800$, $1015$ 
respectively) similar to the $\barmu=635$ case reported in 
Fig.~\ref{fig:fig1}d,e, though here we observe a larger oscillation amplitude 
characterized by $\nonOmegaamp$; 
and a steady spinning motion 
(Fig.~\ref{fig:fig2}c,d, where $\barmu=1020$, $2000$ respectively), resembling 
a 
QR particle towing a passively bent filament that breaks the mirror symmetry  
about the filament centerline;
a stationary state when $\barmu$ is below a critical value, 
where the  object is stationary ($\nonOmega=0$) and
 possesses mirror symmetry. 
The three elasticity-dependent states are identified with a dashed 
line (stationary), triangles (undulatory) and diamonds (spinning) in 
Fig.~\ref{fig:fig2}e, which represent
 the bifurcation diagram of a
 one-parameter ($\barmu$) dynamical system: 
the stationary state is a symmetric fixed-point solution, which transits 
through 
a supercritical Hopf bifurcation~\cite{steven1994nonlinear} at $\barmucrione 
\approx 625$
 to a limit-cycle solution corresponding to the undulating state. 
 This periodic solution jumps, 
via a secondary  bifurcation at 
$\barmucritwo \approx 1017$,
 to another asymmetric fixed-point solution representing the spinning state.
 The Hopf bifurcation is confirmed by the quadratic variation of 
$\nonOmegaamp$ in $\barmu$ in the vicinity of 
$\barmucrione$ shown in Fig.~\ref{fig:fig2}b (linear dependence of 
$\lp\nonOmegaamp\rp^2$ on $\barmu$). It is worth-noting that the 
bifurcation 
diagram featured with these two bifurcations remains unchanged when the 
electric field $1 < \barE < \bar{\mathcal{E}}^{\mathrm{cri}}$, where 
$\bar{\mathcal{E}}^{\mathrm{cri}}$ corresponds to the critical field above which 
the particle with a rigid filament ($\barmu = 0$) undergoes QR; when 
$\barE>\bar{\mathcal{E}}^{\mathrm{cri}}$, the object spins steadily
regardless of $\barmu$.

The composite object achieves  self-oscillatory propulsion only in the 
undulating regime $\barmu\in \lp \barmucrione, \barmucritwo \rp$,
attaining zero net locomotion when $\barmu \rightarrow \barmucrione+$ 
and $\barmu \rightarrow \barmucritwo-$.
We expect its propulsive performance to exhibit a non-monotonic dependence 
on $\barmu$ and  peaks at an optimal EEV 
parameter $\barmuopt$.  
We quantify the performance by the translational velocity $\Utran$ of the 
swimmer along its 
effective straight path connecting the most convex points on the 
wavelike
trajectory (Fig.~\ref{fig:fig3}a). The trajectory shape  
depends on $\barmu$: for the stiffest filament 
$\barmu=650$, it matches a sinusoidal wave with a high frequency,
almost preserving 
fore-aft temporal symmetry. Conversely, when $\barmu=825$, 
the wavy trajectory is
characterized by 
a larger amplitude and lower frequency. For the most floppy case 
shown $\barmu=950$, 
the trajectory is significantly
 coiled, exhibiting a
pronounced fore-aft asymmetry. 
Consequently, the swimmer's backward movement is comparable to the 
forward movement, leading
to a nearly reciprocal motion.
The increasing coiled trajectory for $\barmu$ is closely linked to the more 
deflected filament shown 
in  Fig.~\ref{fig:fig3}b. 
Fig.~\ref{fig:fig3}c  confirms our anticipation of the non-monotonically varying 
$\Utran$ with  peak value of $\Utranopt \approx 6\times 10^{-3}$ at $\barmuopt 
\approx 825$. This velocity $\Utranopt$ lies in the range  $\lp1, 
15\rp\times 10^{-3}$ of the dimensionless speed of a magnetically-driven 
flexible flagellum~\cite{dreyfus2005microscopic}, implying the 
reasonable efficiency of this self-oscillatory propulsion mechanism.
\ADD{We notice that $\Utran$ oscillates with $\barmu$ when $\barmu \rightarrow 
\barmucritwo$. We do not attempt to unravel this peculiar variation here,
keeping in mind that the main focus of the current work is on 
engineering self-oscillation to achieve various functionalities such as 
locomotion. A thorough analysis 
on the propulsive features of the microrobot will be conducted in future work.}


\begin{figure*}[!ht]
\begin{center}
\hspace{0em}\includegraphics[width=1\textwidth]
{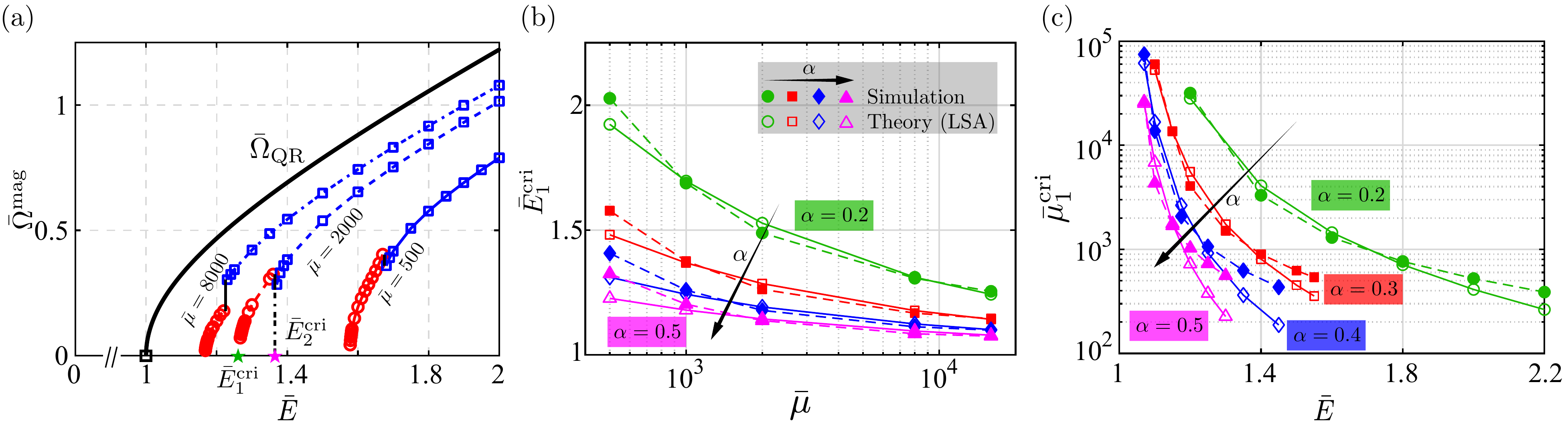}
\end{center}
\caption{(a) Bifurcation diagram 
considering $\barE$ as the control parameter for different $\barmu$ values, 
where
$\barEcrione$ and $\barEcritwo$ indicate the emergence of the Hopf and the
 secondary bifurcations, respectively. $\bar{\Omega}_{\mathrm{QR}}$ denotes 
the 
original  QR (no filament, $\barmu \rightarrow \infty$) velocity. (b)
Critical electric field strength $\barEcrione$ versus $\barmu$,  and (c) 
Critical EEV number $\barmucrione$ versus $\barE$, for different size ratio 
$\alpha$; filled and hollow symbols denote, 
respectively, the numerical and theoretical (LSA) 
predictions.}
\label{fig:fig4}
\end{figure*} 

By varying $\barE$, we present a
bifurcation diagram in Fig.~\ref{fig:fig4}a for the composite object with 
different $\barmu$ values. The
diagram shares the same feature with Fig.~\ref{fig:fig2}e 
considering $\barmu$ as the control parameter: a Hopf and a secondary 
bifurcation
occur at $\barEcrione$ and $\barEcritwo$, respectively. 
$\barE=1$ (hollow square) indicates the pitchfork bifurcation resulting in 
the original QR instability.
This graph has highlighted the role of the filament  
in transforming the pitchfork bifurcation into the Hopf bifurcation that leads 
to self-oscillation. 
\ADD{We further conduct a linear stability analysis 
(LSA)~\cite{zhu2018elasto-long2} around an equilibrium base solution $[\theta, 
\nP_Q, \nP_3] = 
[\hat{\theta}, -\kappa \bareta \barE \sin{\hat{\theta}}, -\kappa 
\bareta \barE \cos{\hat{\theta}}]$ of 
Equ.~\ref{eq:non-euler} ($\hat{\theta}$
can be an arbitrary value without loss of generality), 
when the composite system is stationary and the filament is undeformed.
Without the filament, $\nTfp_N=0$, the LSA indeed predicts a 
critical electrical field of $\barE=1$ corresponding to that of 
the original QR instability.  In the presence of filament, realizing that the 
filament 
undergoes weak deformation near the onset of instability,
we are able to model the elastic torque $\nTfp_N$ following  
Refs.~\cite{wiggins1998flexive,wiggins1998trapping}.
The  critical electric field $\barEcrione$  predicted by
LSA above which self-oscillatory 
instability occurs is shown as a function of $\barmu$ and $\alpha$} 
(Fig.~\ref{fig:fig4}b), so as
the critical EEV number $\barmucrione$  versus $\barE$ and $\alpha$ 
(Fig.~\ref{fig:fig4}c). These theoretical predictions agree well with their 
numerical counterparts, especially when $\barmu$ is large.

In this work, we have uncovered an EEH 
instability and  demonstrated a strategy for engineering self-oscillations
based on a steady, uniform electric field. This idea is illustrated by 
driving the motion of a dielectric particle 
connected to an 
elastic filament. 
By tuning the filament elasticity and electric field strength, the object
achieves propulsion enabled by the dual functionalities of the filament: 
manipulating the bifurcation through its elasto-viscous response, 
which causes the particle to oscillate; providing thrust by wiggling motion 
actuated by the oscillating particle.  
Besides offering the possibility as a swimming
microrobot, the object can also transform into a stationary, soft obstacle or 
spinner when the electric field strength is tuned, respectively,  below 
or above the critical
values we have identified.

The key idea we have recognized is to introduce an
elastic element to trigger the Hopf bifurcation 
and consequently the self-oscillatory instability.
Therefore, this might be one path for engineering biomimetic
oscillatory processes using a time-independent power source. More 
generally, our results imply the potential for
incorporating elastic media in other unstable systems to manipulate
and diversify the bifurcations~\cite{chen2000bifurcation}, which possibly 
can be employed for different functionalities. This concept is different from, 
but 
complementary to, taming a structure's mechanical failures to 
achieve functionalities~\cite{reis2018mechanics}.
We believe that our ideas offer opportunities to develop a new  
generation of soft, reconfigurable machines that can morph and adapt to the 
environment. Experiments to test and further explore the EEH instability 
introduced in this work are in progress.

We thank Drs. Y. Man, L. Li and G. Balestra, and Profs. O. S. Pak, B. 
Rallabandi, E. Nazockdast, Y. N. Young  and F. Gallaire for
useful discussions. Prof. T. G\"{o}tz is acknowledged for sharing us with his 
Phd thesis. L.Z. thanks 
the Swedish Research Council for a VR International Postdoc Grant (2015-06334). 
We thank the NSF for support
via the Princeton University Material Research Science and Engineering Center 
(DMR-1420541). The computer time was
provided by SNIC (Swedish National Infrastructure for Computing). 


\begin{thebibliography}{25}
\bibitem [{bri()}]{bridge}%
  \BibitemOpen
\bibfield  {journal} {  }\href@noop {} {\bibinfo  {journal} {This phenomenon is
  mischaracterized by some introductory physics texts, where it is attributed
  to forced resonance}\ }\BibitemShut {NoStop}%
\end{thebibliography}

%

\end{document}